# Real-Space Selection Rule for Moiré Exciton-Phonon Coupling

Yixuan Liu[1,2#], Shiyu Peng[1,2#], Yifan Wang[1,2], Peiyao Qin[1,2], Wang Yao[3], Chen Hu[1,2*]

*[1] Department of Physics, The Chinese University of Hong Kong, Hong Kong, China*

*[2] State Key Laboratory of Quantum Information Technologies and Materials, The Chinese University of Hong Kong, Hong Kong, China*

*[3] New Cornerstone Science Lab, Department of Physics, The University of Hong Kong, Hong Kong, China*

# These authors contributed equally

* Email: chenhu@cuhk.edu.hk

**Abstract:**

Exciton-phonon coupling (ExPC) governs energy relaxation and exciton dynamics in photoexcited quantum materials. Recent advances have revealed rich and tunable excitonic phenomena in moiré superlattices, yet the microscopic mechanisms underlying the coupling between moiré excitons and moiré phonons remain elusive. Here, we develop a real-space representation for moiré ExPC based on a unified theoretical framework combining the many-body Bethe-Salpeter equation with phonon perturbation theory. Furthermore, we uncover a general real-space selection rule: the spatial matching of a moiré exciton's electron and hole densities with the moiré-phonon-induced potential perturbations determines their individual coupling contributions, while coherent interference between the electron and hole contributions controls the total ExPC. We demonstrate this rule in prototypical $WSe_2/WS_2$ moiré heterobilayers, revealing pronounced exciton- and phonon-mode-dependent coupling, including nearly complete suppression of ExPC through destructive electron-hole interference. These findings establish the internal electron-hole spatial configuration of moiré excitons as a key degree of freedom governing the strength and mode selectivity of their coupling to moiré phonons, opening opportunities to engineer ExPC through spatially tailored moiré potentials.

## Main text:

Stacking two atomically thin crystals with a slight rotational or lattice mismatch creates an emergent periodicity on a length scale far exceeding that of the underlying atomic lattices. The resulting long-wavelength moiré potential profoundly reshapes electronic states, giving rise to a broad range of emergent quantum phenomena, including correlated insulators (1-8), unconventional superconductivity (7, 9-13), magnetism (14-20), generalized Wigner crystals (21-23), and topological phases (24-31). Upon photoexcitation, the interplay between the moiré potential and electron-hole Coulomb attraction in transition-metal dichalcogenide moiré superlattices produces moiré excitons with rich and highly nontrivial internal real-space structures (32-37). Recent advances have further demonstrated that these internal electron-hole configurations can be tuned through the moiré potential, establishing their spatial structure as a distinctive and controllable degree of freedom (36, 38).

The moiré potential is highly sensitive to local atomic registry and lattice relaxation(32, 36, 37). Competition between stacking energetics and intralayer elasticity drives structural reconstruction, producing displacement and strain textures across the moiré superlattice that reshape the local electronic landscape (39-42). The same lattice degrees of freedom also support low-energy moiré phonons, which correspond to dynamical fluctuations about the reconstructed equilibrium structure and modulate the electron and hole moiré potentials (43-48). Structural reconstruction and moiré phonons are naturally unified as the equilibrium and dynamical components of the same lattice displacement field (49). Exciton-phonon coupling (ExPC) plays a central role in exciton relaxation, dephasing, linewidth broadening, and nonequilibrium dynamics (50-53). When a moiré exciton couples to a moiré phonon, both the excitonic electron-hole distributions and the phonon-induced potential perturbations can exhibit diverse real-space structures across the moiré supercell. Their spatial interplay therefore emerges as a key to understanding moiré ExPC. Moreover, beyond the coupling of individual electronic states, ExPC involves both electron-phonon and hole-phonon scattering channels, whose contributions combine coherently (54-56). However, the microscopic real-space principle governing the strength and mode selectivity of the moiré ExPC remains elusive.

In this work, we develop a real-space representation of moiré ExPC based on a unified theoretical framework combining the many-body Bethe-Salpeter equation with phonon perturbation theory. We uncover a real-space selection rule: spatial matching between the excitonic electron and hole distributions and the corresponding phonon-induced moiré potentials determines the magnitude and phase of their individual coupling contributions, while coherent interference between these contributions controls the total ExPC. We demonstrate this rule in prototypical $WSe_2/WS_2$ moiré superlattices by examining two types of excitons with distinct electron-hole spatial configurations, i.e. charge-transfer (CT) and Wannier-type moiré excitons, coupled to two dominant moiré phonon modes with contrasting spatial patterns. Our results establish spatial matching as a microscopic principle governing the exciton- and phonon-mode selectivity of moiré ExPC.

Fig. 1 illustrates the real-space mechanism of moiré ExPC. The microscopic connection between structural reconstruction and moiré phonons can be descried through the lattice displacement field (44, 45),

$$\mathbf{u}(\mathbf{r}) = \mathbf{u}_0(\mathbf{r}) + \delta\mathbf{u}_\mu(\mathbf{r}), \tag{1}$$

where $\mathbf{u}_0(\mathbf{r})$ describes the equilibrium structural reconstruction and $\delta\mathbf{u}_\mu(\mathbf{r})$ denotes the dynamical displacement associated with phonon mode μ. Through a common displacement-to-potential mapping (as detailed below), the equilibrium displacement $\mathbf{u}_0(\mathbf{r})$ determines the static electron and hole moiré potentials $V_0^{(e/h)}(\mathbf{r})$, whereas the dynamical displacement $\delta\mathbf{u}_\mu(\mathbf{r})$ generates the corresponding phonon-induced potential perturbations $\delta V_\mu^{(e/h)}(\mathbf{r})$ (49). As illustrated in Fig. 1(b), the equilibrium moiré potentials $V_0^{(e/h)}(\mathbf{r})$ localize electrons and holes in characteristic stacking regions, while moiré phonons dynamically modulate these potential landscapes. The resulting moiré electron(hole)-phonon coupling depends sensitively on the spatial matching between the electronic state distributions and the corresponding phonon-induced perturbations $\delta V_\mu^{(e/h)}(\mathbf{r})$. For moiré excitons, the spatial distributions of electron and hole are determined by the interplay between their equilibrium moiré potentials and their mutual Coulomb interaction (32,26). The electron and hole components contribute coherently to the total phonon-induced scattering amplitude (54-56). Beyond the spatial matching within each channel, the total ExPC also depends on the relative magnitudes and phases of the two contributions. Their constructive or destructive interference can strongly enhance or suppress the total coupling.

We consider a rotationally aligned $WSe_2/WS_2$ heterobilayer, a prototypical moiré semiconductor that has attracted considerable attention for hosting rich moiré exciton physics (32, 35, 37). The approximately 4% lattice mismatch produces a moiré superlattice with a period of about 8.3 nm. As illustrated in Fig. 2(a), the local atomic registry varies continuously across the moiré supercell, connecting the three high-symmetry stacking configurations AA, $B^{Se/W}$, and $B^{W/S}$. The $WSe_2/WS_2$ heterobilayer has a type-II band alignment, while its low-energy optical response is dominated by intralayer excitons in $WSe_2$ (32, 37). We therefore focus below on the $WSe_2$ intralayer excitonic states and their coupling with moiré phonons. Owing to the different local stacking energies (see Fig. S1(b) in the Supporting Information), the moiré superlattice undergoes substantial structural reconstruction. The competition between stacking-dependent interlayer interactions and intralayer elasticity enlarges the energetically favorable stacking domains $B^{Se/W}$ and $B^{W/S}$ at the expense of the higher-energy AA regions (32, 37). We describe this reconstruction within a continuum model that includes both contributions and obtain the relaxed structure by minimizing the total energy with respect to the in-plane atomic displacement field (44, 45) (see Supporting Information for more details). Figure 2(b) shows the resulting displacement field $\mathbf{u}_0(\mathbf{r})$ in the $WSe_2$ layer, which directly visualizes the reconstruction of the moiré lattice. The spatial variation of $\mathbf{u}_0(\mathbf{r})$ generates an in-plane strain field described by the symmetric strain tensor: $u_{ij}^0 = \frac{1}{2}(\partial_i u_j^0 + \partial_j u_i^0)$, where $i, j = x, y$ (49, 57-59). Figure 2(c) shows the isotropic strain component $u_{xx}^0 + u_{yy}^0$, together with the anisotropic components $(u_{xx}^0 - u_{yy}^0, -2u_{xy}^0)$. The strain distribution closely follows the reconstructed stacking landscape. In the $WSe_2$ layer, structural reconstruction leads to local compression in the low-energy stacking regions $B^{Se/W}$ and $B^{W/S}$, while tensile strain is concentrated near the AA regions. By contrast, the anisotropic strain components are strongest along the domain walls separating neighboring stacking domains. As shown below, these reconstruction-induced strain textures provide the microscopic origin of the spatially varying moiré potentials.

To describe moiré exciton-phonon coupling microscopically, it is essential to establish a direct connection between lattice distortions and the moiré potential. Here, we develop such a mapping and identify the microscopic origin of its low-energy moiré potential. Because interlayer hybridization between the $WSe_2$ and $WS_2$ bands is relatively weak (32), the moiré electronic landscape can be described directly in terms of the strain field generated by the lattice displacement $\mathbf{u}(\mathbf{r})$. We find that the resulting moiré potential experienced by electrons and holes can be accurately decomposed into scalar deformation-potential and piezoelectric contributions:

$$V^{(e/h)}[\mathbf{u}](\mathbf{r}) = V_s^{(e/h)}[\mathbf{u}](\mathbf{r}) + V_p^{(e/h)}[\mathbf{u}](\mathbf{r}), \tag{2}$$

The scalar contribution $V_s^{(e/h)}(\mathbf{r})$ is governed by the isotropic strain $u_{xx} + u_{yy}$ (57, 58). In addition, because monolayer TMDs lack inversion symmetry, lattice distortion induces an in-plane piezoelectric polarization (61). The spatially nonuniform strain field of the moiré superlattice therefore generates a periodic piezoelectric charge density $\rho_p(\mathbf{r}, t) = -\tilde{e}_{11}\nabla \cdot (u_{xx} - u_{yy}, -2u_{xy})$, where $\tilde{e}_{11}$ is the in-plane piezoelectric coefficient that characterizes the linear polarization response to strain (59-61). $\rho_p(\mathbf{r})$ in turn produces the electrostatic potential $V_p^{(e/h)}(\mathbf{r})$ (59, 60). For the equilibrium structural reconstruction $\mathbf{u}_0(\mathbf{r})$, this mapping yields the static moiré potentials $V_0^{(e/h)}(\mathbf{r}) = V_{s,0}^{(e/h)}(\mathbf{r}) + V_{p,0}^{(e/h)}(\mathbf{r})$, as shown in Figs. 2(d) and 2(e). The same displacement-to-potential relation in Eq. (2) also applies to dynamical lattice distortions, such that a phonon displacement $\delta\mathbf{u}_\mu(\mathbf{r})$ generates the corresponding phonon-induced perturbation $\delta V_\mu^{(e/h)}(\mathbf{r})$ as discussed below (Fig. 4). Details of the corresponding continuum Hamiltonian, strain-dependent potential terms, and moiré potential parameters are given in Methods and Supporting Information.

Figures 2(d) and 2(e) reveal that the reconstruction-induced equilibrium electron and hole moiré potentials $V_0^{(e/h)}(\mathbf{r})$ emerge from markedly different spatial combinations of these two contributions. For electrons, $V_{s,0}^{(e)}(\mathbf{r})$ and $V_{p,0}^{(e)}(\mathbf{r})$ have comparable magnitudes but distinct spatial profiles. The tensile strain around AA produces a large negative scalar deformation potential, making AA a favorable trapping region for electrons. The anisotropic strain, concentrated primarily near the domain walls, generates a strongly modulated piezoelectric potential around the two Bernal stackings. At $B^{Se/W}$, the scalar and piezoelectric contributions combine to raise the electron potential, whereas at $B^{W/S}$ they partially compensate. The equilibrium electron moiré landscape $V_0^{(e)}(\mathbf{r})$ therefore results from the spatial interplay of the two contributions rather than either term alone. For holes, the scalar deformation contribution is negligible and the equilibrium moiré potential $V_0^{(h)}(\mathbf{r})$ is dominated by the piezoelectric term $V_{p,0}^{(h)}(\mathbf{r})$, with its minimum located at the $B^{Se/W}$ stacking. The electron and hole potential minima consequently occur at different local registries, naturally favoring spatially separated electron-hole configurations and the formation of intralayer charge-transfer excitons discussed below. Remarkably, our computed low-energy electronic structure obtained from the equilibrium strain-based moiré potential agrees quantitatively with that from large-scale ab initio calculations of the fully reconstructed superlattice (32) (see Fig. S2(a,b) in the Supporting Information), validating the accuracy of the established mapping from the displacement field to the moiré electronic landscape.

With the microscopic equilibrium states established, we next determine the excitonic states by solving the Bethe-Salpeter equation (BSE) (62, 63),

$$(E_{c\mathbf{k}} - E_{v\mathbf{k}})A_{vc\mathbf{k}}^S + \sum_{v'c'\mathbf{k}'} K_{vc\mathbf{k},v'c'\mathbf{k}'}\, A_{v'c'\mathbf{k}'}^S = \Omega_S A_{vc\mathbf{k}}^S, \tag{3}$$

where $E_{c\mathbf{k}}$ and $E_{v\mathbf{k}}$ are the conduction and valence band energies, $\Omega_S$ is the exciton energy, and $A_{vc\mathbf{k}}^S$ is the exciton eigenvector. The interaction kernel $K = V - W$ contains the bare exchange interaction $V$ and the screened direct Coulomb attraction W (see Methods and Supporting Information for details). The calculated absorbance spectrum in Fig. 3(a) exhibits three prominent moiré-exciton resonances, labeled I to III. These moiré resonances possess distinctly different internal real-space structures. To visualize them, we compute the electron and hole densities from the exciton wavefunction $\chi_S(\mathbf{r}_e, \mathbf{r}_h)$: $\rho_S^e(\mathbf{r}_e) = \int |\chi_S(\mathbf{r}_e, \mathbf{r}_h)|^2 d\mathbf{r}_h$ and $\rho_S^h(\mathbf{r}_h) = \int |\chi_S(\mathbf{r}_e, \mathbf{r}_h)|^2 d\mathbf{r}_e$. As shown in Fig. 3(b), exciton I is a moiré-modulated Wannier-like exciton with the electron and hole densities concentrated around the AA region. In contrast, exciton III exhibits pronounced intralayer charge-transfer (CT) character, with the electron localized near AA and the hole near $B^{Se/W}$ region. Exciton II has a mixed character (32, 37). Both the excitonic spectrum and the real-space characters are consistent with recent experiments and large-scale ab initio GW-BSE calculations (32, 37) (see Fig. S2 in Supporting Information).

Next, we will examine the coupling of these moiré excitons to moiré phonons. Within the BSE-based ExPC framework, the coupling of an exciton state S to a phonon mode μ is given by the coherent sum of electron-phonon and hole-phonon contributions (54-56):

$$G_{S\mu} = G_{S\mu}^e + G_{S\mu}^h \,. \tag{4}$$

Their microscopic expressions are $G_{S\mu}^e = \sum_{\mathbf{k}vcc'} A_{vc'\mathbf{k}}^{S*} A_{vc\mathbf{k}}^S g_{c'c\mu}(\mathbf{k})$ and $G_{S\mu}^h = -\sum_{\mathbf{k}cvv'} A_{vc\mathbf{k}}^{S*} A_{v'c\mathbf{k}}^S g_{v'v\mu}(\mathbf{k})$ (54-56), where $g_{mn\mu}(\mathbf{k}) = \langle m\mathbf{k}|\delta H_\mu|n\mathbf{k}\rangle$ denotes the electron-phonon matrix element of phonon mode μ. $\delta H_\mu$ is the first-order variation of the electronic Hamiltonian induced by the phonon displacement $\delta\mathbf{u}_\mu(\mathbf{r})$ (49, 64). The two terms describe phonon scattering through the electron and hole components of the exciton wavefunction, respectively. Computational details of the moiré phonon and ExPC calculations are provided in the Methods and Supporting Information. Figs. 3(c) and 3(d) show the calculated ExPC strengths for the low-energy moiré phonon branches of the CT and Wannier excitons, respectively. The calculated ExPC exhibits two prominent features: first, the CT exciton couples substantially more strongly than the Wannier exciton across the low-energy moiré-phonon spectrum; second, the interaction is dominated by two phonon modes, labeled A and B, with energies of 2.54 and 2.71 meV, respectively. This pronounced dependence on both exciton character and phonon mode points to a highly selective coupling mechanism.

To elucidate the origin of the contrasting ExPC, we examine the electron-phonon contribution $G_{S\mu}^e$ and the hole-phonon contribution $G_{S\mu}^h$ introduced in Eq. (4). As shown in Figs. 3(e) and 3(f), these two contributions exhibit qualitatively different behavior for the CT and Wannier excitons. For the CT exciton, $G_{CT}^e$ and $G_{CT}^h$ are both sizable for modes A and B and have the same sign, resulting in constructive interference and a strongly enhanced total coupling. For the Wannier exciton, the electron and hole contributions instead have opposite signs. This contrast is most distinguished for mode A. The magnitudes of $G_{W,A}^e$ and $G_{W,A}^h$ are 0.24 and 0.22 meV, respectively, comparable to the corresponding values of 0.28 and 0.21 meV for the CT exciton ($G_{CT,A}^e$ and $G_{CT,A}^h$). However, the two contributions nearly cancel in the Wannier case but add constructively in the CT case, yielding total couplings of 0.02 and 0.49 meV, respectively, a difference of more than one order of magnitude. These results demonstrate how coherent interference between the electron and hole contributions can either enhance or suppress moiré ExPC, as exemplified by the CT and Wannier excitons considered here, respectively.

The fundamental distinction between Wannier and CT excitons lies in their internal real-space electron-hole configurations. A real-space formulation therefore provides a natural framework for understanding the contrasting ExPC revealed above. Starting from the microscopic BSE exciton-phonon matrix element, we derive an explicit real-space representation:

$$G_{S\mu}^{e/h} = \int \rho_S^{e/h}(\mathbf{r}) \delta V_\mu^{(e/h)}(\mathbf{r})\, d\mathbf{r}, \tag{5}$$

where $\rho_S^{e/h}(\mathbf{r})$ are the electron and hole density distributions of exciton S, and $\delta V_\mu^{(e/h)}(\mathbf{r})$ are the corresponding phonon-induced moiré-potential perturbations of mode $\mu$. This expression establishes a direct microscopic connection between the internal real-space structure of a moiré exciton and its coupling to lattice motion. This real-space representation is further validated against the full ExPC calculation and yields nearly identical ExPC matrix elements (see Fig. S3 in Supporting Information). The magnitude and sign of each electron or hole contribution are thus determined by the density-weighted spatial overlap with the corresponding phonon-induced moiré potential.

Based on the Eq. (5), Figs. 4(a) and 4(b) further reveal the spatial origin of the contrasting moiré ExPC for the dominant phonon modes A and B. These two modes exhibit distinct displacement patterns and consequently generate different spatial modulations of the electron and hole potential landscapes. For mode A, the isotropic strain modulation is concentrated around the two Bernal stacking regions with opposite signs, whereas mode B exhibits opposite-sign strain modulations around AA and the two Bernal stacking regions. The resulting phonon-induced potential perturbations $\delta V_\mu^{(e)}(\mathbf{r})$ and $\delta V_\mu^{(h)}(\mathbf{r})$ vary strongly across the moiré supercell, with phonon-mode-dependent spatial profiles and sign changes between different stacking regions. A detailed analysis of these phonon-induced potentials is provided in the Supporting Information.

For the CT exciton, the electron (hole) density is predominantly localized at the AA ($B^{Se/W}$) stacking region [Fig. 3(b)]. As shown in Figs. 4(a) and 4(b), both electron and hole components are localized in regions with large phonon-induced potential perturbations for modes A and B ($\delta V_{A/B}^{(e)}(\mathbf{r})$ and $\delta V_{A/B}^{(h)}(\mathbf{r})$). This strong spatial matching gives rise to sizable electron-phonon and hole-phonon contributions, $G_{CT}^{e}$ and $G_{CT}^{h}$ for both modes [Fig. 3(e)]. Moreover, $\delta V_{A/B}^{(e)}(\mathbf{r})$ and $\delta V_{A/B}^{(h)}(\mathbf{r})$ have the same sign in the regions where the respective electron/hole components are localized, leading to constructive interference and strongly enhanced ExPC [Fig. 3(c)]. For the Wannier exciton, both electron and hole densities are localized at the AA stacking region. For mode B [Fig. 4(b)], the electron (hole) density overlaps with a strong (weak) $\delta V_{B}^{(e)}(\mathbf{r})$ ($\delta V_{B}^{(h)}(\mathbf{r})$). This spatial matching (mismatch) of the electron (hole) channel yields a sizeable (negligible) coupling contribution $G_{W,B}^{e}$ ($G_{W,B}^{h}$) in Fig. 3(f). For mode A, by contrast, both electron and hole channels exhibit strong spatial matching, producing sizable individual contributions $G_{W,A}^{e}$ and $G_{W,A}^{h}$ [Fig. 3(f)]. However, $\delta V_{A}^{(e)}(\mathbf{r})$ and $\delta V_{A}^{(h)}(\mathbf{r})$ have opposite signs at AA stacking (where both electron and hole are located), leading to nearly complete destructive interference and a strongly suppressed total ExPC [Fig. 3(d)]. Together, these results establish a real-space selection rule for moiré ExPC: strong coupling is favored when both electron-phonon and hole-phonon channels exhibit substantial spatial matching and their contributions interfere constructively, whereas spatial mismatch or destructive interference can strongly suppress it.

In conclusion, we have uncovered a real-space selection rule for moiré ExPC: the magnitude and sign of the individual electron and hole coupling contributions are determined by their spatial overlaps with the corresponding phonon-induced potentials, while their coherent interference controls the total ExPC. For reconstructed $WSe_2/WS_2$ superlattices, this rule explains the contrasting couplings of moiré charge-transfer and Wannier excitons to two dominant phonon modes with distinct spatial patterns. More broadly, our findings establish the internal electron-hole spatial configuration of moiré excitons as a key degree of freedom governing ExPC in moiré superlattices. The ability to tailor these spatial configurations through moiré potentials thus offers a promising route to controlling phonon-mediated relaxation, coherence, and nonequilibrium dynamics. The real-space framework developed here provides a microscopic foundation for understanding and engineering exciton-phonon interactions in moiré semiconductors. Moreover, although this work focuses on moiré excitons, the underlying real-space matching principle can also be extended to moiré electron-phonon coupling, offering new insights into phonon-mediated superconductivity and other phonon-driven quantum phenomena in moiré materials.

## Methods:

### Lattice reconstruction and moiré phonon calculations

We describe the structural reconstruction and low-energy in-plane moiré phonons of the rotationally aligned $WSe_2/WS_2$ heterobilayer within a continuum elasticity framework (44, 45). The monolayer lattice constants are 3.32 and 3.19 Å for $WSe_2$ and $WS_2$, respectively, corresponding to a moiré period of about 8.3 nm (32, 37). The continuum energy includes intralayer elastic energy $U_e$ and a lowest-harmonic, stacking-dependent interlayer binding energy $U_b$. The equilibrium displacement fields are obtained by iteratively solving the continuum Euler-Lagrange equations in momentum space. Moiré phonons are calculated by linearizing these equations about the reconstructed equilibrium structure and diagonalizing the resulting dynamical matrix. Each phonon mode $\mu$ is characterized by an eigenfrequency $\omega_\mu$ and a displacement field $\delta\mathbf{u}_\mu(\mathbf{r})$, with the corresponding time dependence given by $\delta\mathbf{u}_\mu(\mathbf{r},t) = e^{-i\omega_\mu t}\delta\mathbf{u}_\mu(\mathbf{r})$. Throughout this work, $\delta\mathbf{u}_\mu(\mathbf{r})$ and $\delta V_\mu^{e/h}(\mathbf{r})$ denote the spatial profiles of the phonon displacement and the corresponding induced potential perturbations, respectively. Further details are provided in Supporting Information.

### Electronic continuum models and moiré-potential calculations

The low-energy $WSe_2$ electronic states are described by a two-band continuum Hamiltonian $H$ parametrized against first-principles calculations (29, 36, 65). For the reconstructed equilibrium structure, the Hamiltonian takes the form $H_0 = \hbar v_F\left(\Delta\mathbf{k} + \frac{\mathbf{e}}{\hbar}\mathbf{A}_0\right)\cdot\boldsymbol{\sigma} + \frac{E_g}{2}\sigma_z + V_0(\mathbf{r})$ where $V_0(\mathbf{r})$ denotes the equilibrium strain-induced moiré potential shown in Fig. 2(d) and Fig. 2(e), and $\mathbf{A}_0$ is the equilibrium pseudovector field. The Hamiltonian is diagonalized in momentum space to obtain the moiré band energies and wavefunctions, which provide a single-particle basis for the excitonic and electron-phonon calculations. Details of the continuum Hamiltonian and moiré-potential parameters are provided in Supporting Information.

### Moiré Exciton Calculations

Intralayer moiré excitons are obtained by solving the BSE [Eq. (3)] in the basis of electron-hole transitions constructed from the continuum-model states (36, 65). The interaction kernel includes the bare exchange interaction V and the screened direct Coulomb attraction W. The optical absorbance is evaluated from the excitonic eigenstates and velocity-matrix elements. The real-space exciton wavefunctions are calculated by: $\chi_S(\mathbf{r}_e,\mathbf{r}_h) = \sum_{cvk} A^S_{\mathbf{vck}}\psi^*_{\mathbf{vk}}(\mathbf{r_h})\psi_{\mathbf{ck}}(\mathbf{r_e})$, where $\psi_{\mathbf{vk}}(\mathbf{r_h})$ and $\psi_{\mathbf{ck}}(\mathbf{r_e})$ are moiré valence and conduction states, respectively. Further computational details are provided in Supporting Information.

**Moiré ExPC Calculations**

For each moiré phonon mode, the phonon-induced electronic perturbation is evaluated to first order in the displacement $\delta\mathbf{u}_\mu(\mathbf{r})$ about the reconstructed equilibrium structure (44, 45). The displacement-to-potential mapping used for constructing the equilibrium moiré potential is also applied to obtain the dynamical phonon-induced potential perturbations from the phonon displacement. Electron-phonon matrix elements are evaluated between the moiré eigenstates, and the exciton-phonon matrix elements are then constructed from the electron- and hole-scattering channels weighted by the BSE amplitudes (54-56). Their coherent sum gives the total ExPC matrix element. Computational details are provided in Supporting Information.

**Author contributions:** C.H. conceived the project. Y.L., S.P., and C.H. developed the real-space theory of moiré exciton-phonon coupling, implemented the computational framework, and performed the numerical calculations. Y.L., S.P., Y.W., P.Q., W.Y., and C.H. analyzed the results and contributed to their physical interpretation. Y.L. and C.H. wrote the manuscript with input from all authors.

**Acknowledgments:** This work is supported by the start-up fund and the Direct Grant (No. 4053726) of the Chinese University of Hong Kong, the National Natural Science Foundation of China (No. 12504222), and the 1+1+1 CUHK-CUHK(SZ)-GDSTC Joint Collaboration Fund (GRDP2026-018), and the State Key Laboratory of Quantum Information Technologies and Materials.

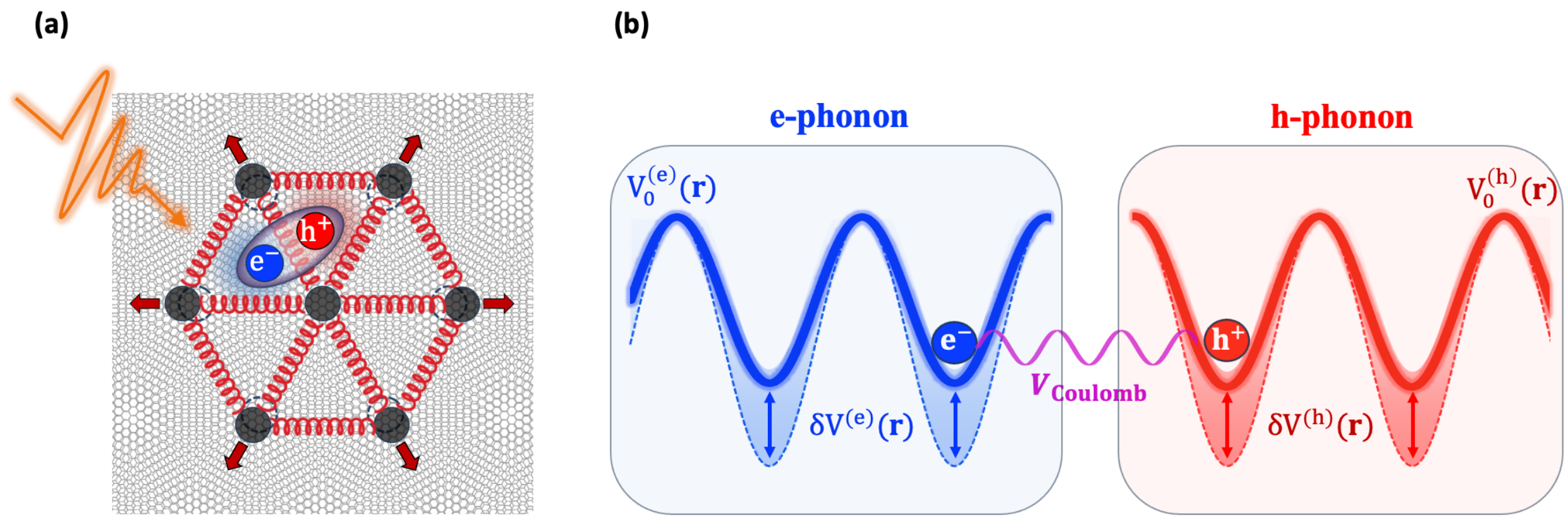

**Fig. 1. Real-space mechanism of moiré exciton-phonon coupling (ExPC). (a)** Schematic illustration of a light-induced moiré exciton interacting with collective lattice vibrations (moiré phonons) in a reconstructed moiré superlattice. **(b)** The total ExPC arises from the coherent superposition of electron-phonon and hole-phonon coupling channels. The electron and hole spatial distributions of the moiré exciton are determined by the interplay between their equilibrium electron-hole moiré potentials $V_0^{(e/h)}(\mathbf{r})$ (solid curves) and the electron-hole Coulomb interaction $V_{coulomb}$. Moiré phonons dynamically modulate these potentials (dashed curves), generating the corresponding phonon-induced perturbations $\delta V^{(e/h)}(\mathbf{r})$, illustrated by the shaded regions between the solid and dashed curves. The electron-phonon and hole-phonon contributions are governed by the spatial matching of the excitonic electron and hole densities with their corresponding phonon-induced potentials.

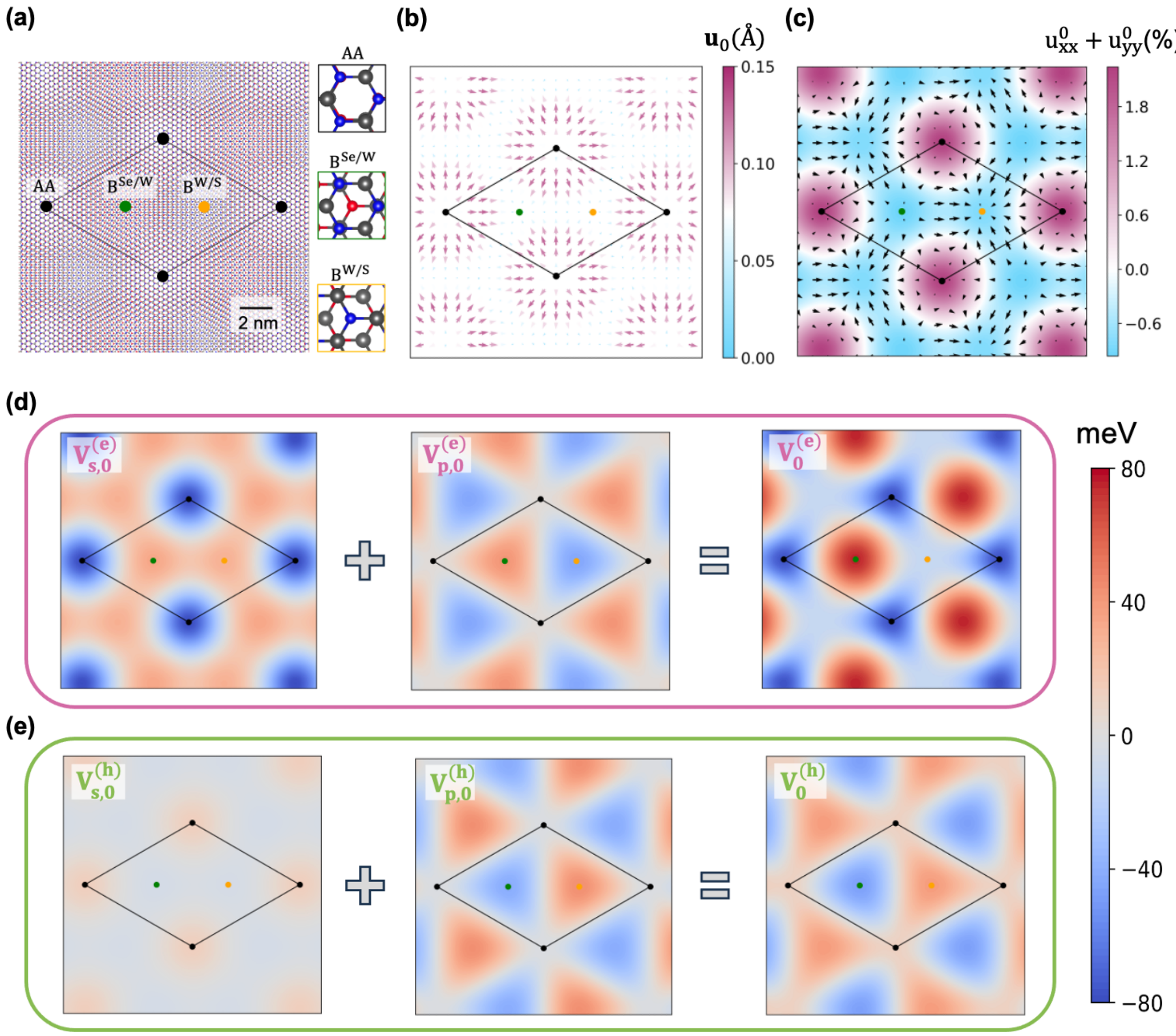

**Fig. 2. Reconstruction-induced equilibrium moiré potentials in $WSe_2/WS_2$ heterobilayer. (a)** Schematic of an R-stacked $WSe_2/WS_2$ moiré superlattice, with the moiré supercell outlined in black and the three high-symmetry stacking sites AA, $B^{Se/W}$, and $B^{W/S}$ marked. **(b)** Equilibrium atomic displacement field $\mathbf{u}_0(\mathbf{r})$ of the reconstructed $WSe_2$ layer. Colors indicate the magnitude of the in-plane displacement, while arrows indicate its direction. **(c)** Isotropic and anisotropic components of the corresponding in-plane strain tensor $u^0_{ij}$. The isotropic component, $u^0_{xx} + u^0_{yy}$, is shown by the color map, while the anisotropic components $(u^0_{xx} - u^0_{yy}\,, -2u^0_{xy})$ are represented by arrows. **(d,e)** Spatial distributions of the resulting equilibrium moiré potential for electrons **(d)** and holes **(e)**. In each panel, the total moiré potential $V_0^{(e/h)}(\mathbf{r})$ (right) is decomposed into the scalar deformation-potential contribution $V_{s,0}^{(e/h)}(\mathbf{r})$ (left) and the piezoelectric potential contribution $V_{p,0}^{(e/h)}(\mathbf{r})$ (middle).

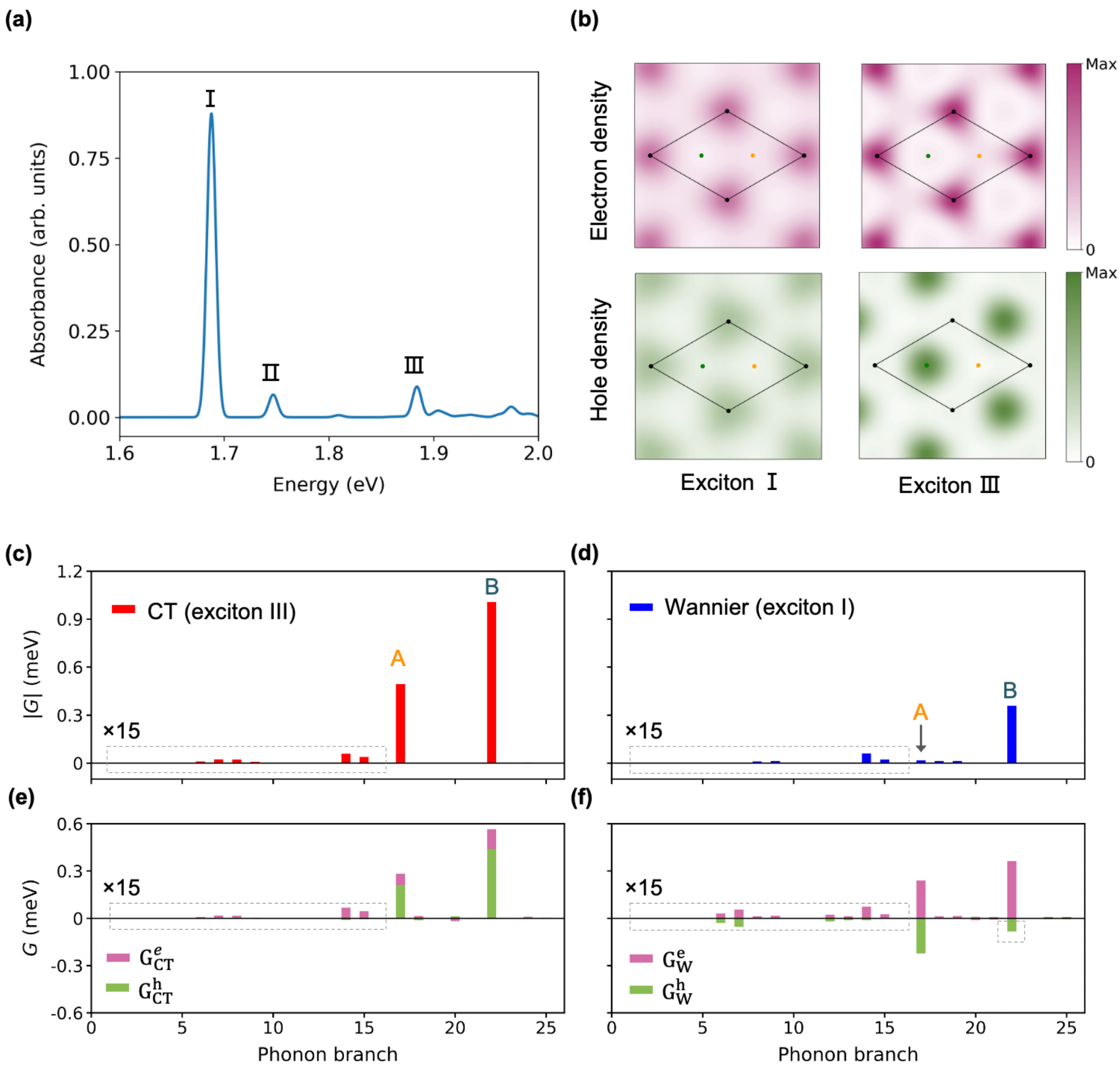

**Fig. 3. Contrasting ExPC of moiré excitons**. **(a)** Calculated absorbance of the moiré superlattice, with the three dominant moiré-exciton resonances labeled I-III. **(b)** Real-space electron and hole densities of excitons I and III. Exciton I is a moiré-modulated Wannier exciton with the electron and hole localized in the same stacking region, whereas exciton III is an intralayer charge-transfer (CT) exciton with the electron and hole localized in different stacking regions. **(c,d)** Magnitudes of the ExPC matrix elements for low-energy moiré phonon modes of the CT exciton III **(c)** and Wannier exciton I **(d)**. The coupling is dominated by phonon modes A and B. **(e,f)** Decomposition of the ExPC into electron-phonon and hole-phonon contributions for the CT **(e)** and Wannier **(f)** excitons. Quantities within the dashed boxes in **(c-f)** are magnified by a factor of 15 for visibility.

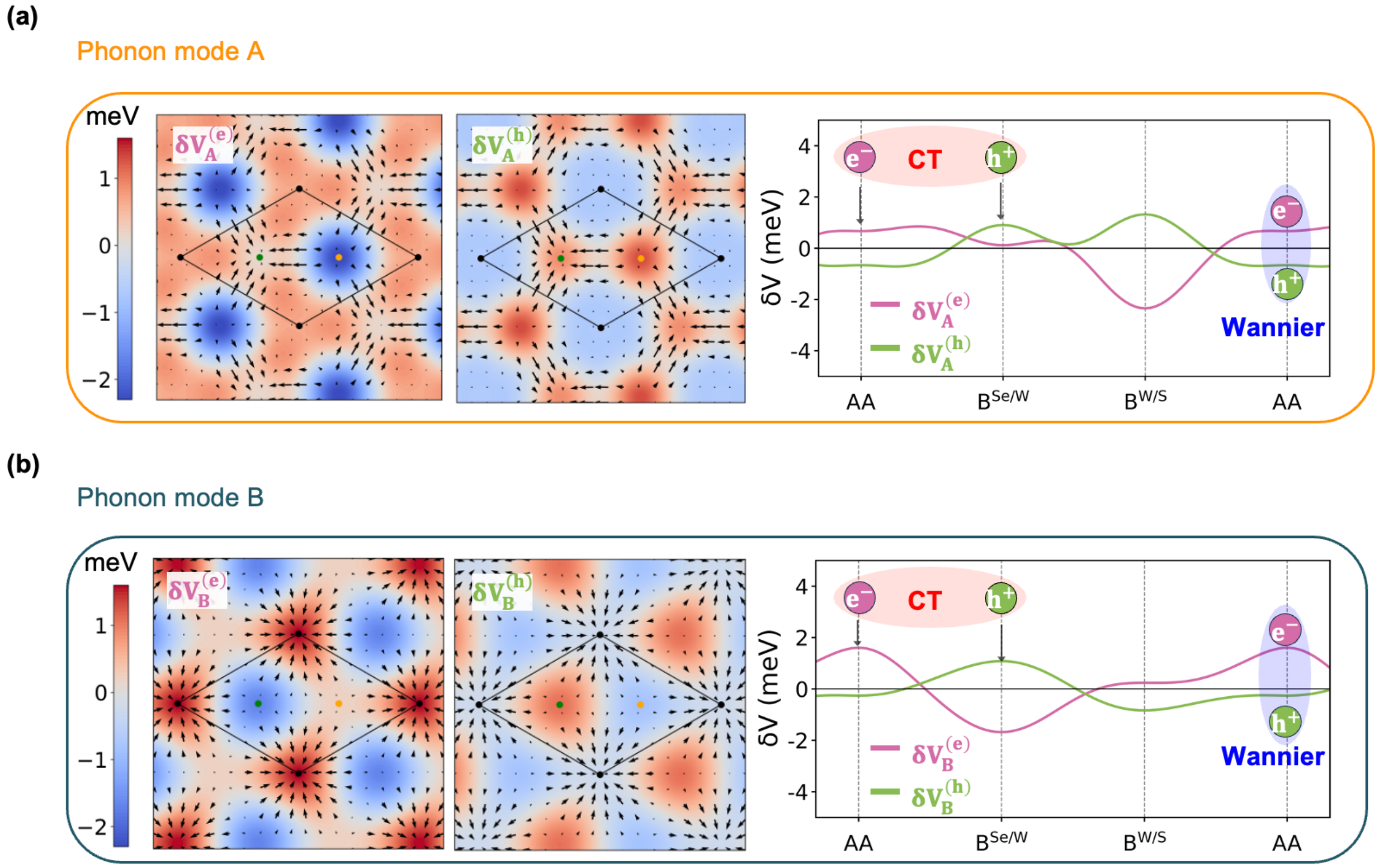

**Fig. 4. Real-space selection rule governing moiré ExPC. (a,b)** Real-space origin of the ExPC for phonon modes A **(a)** and B **(b)**. The left and middle panels show the spatial distributions of the phonon-induced moiré potential perturbations for electrons $\delta V_\mu^{(e)}(\mathbf{r})$ and holes $\delta V_\mu^{(h)}(\mathbf{r})$, respectively, with arrows indicating the corresponding phonon displacement fields. The right panels show line cuts of $\delta V_\mu^{(e)}(\mathbf{r})$ and $\delta V_\mu^{(h)}(\mathbf{r})$ along the indicated high-symmetry path. The electron and hole localization regions of the CT and Wannier excitons are marked to highlight their spatial overlap with the corresponding phonon-induced potential perturbations.